\documentclass[a4paper]{jpconf}

\usepackage{graphicx}
\usepackage{aeguill}
\usepackage{amsmath}
\usepackage{verbatim}

\newcommand{\OO}{{\mathcal O}}

\begin{document}
\title{Doppler-tuned Bragg Spectroscopy of Excited Levels in He-Like Uranium: 
a discussion of the uncertainty contributions}

\author{M.~Trassinelli$^{1}$, A.~Kumar$^{2,3}$, H.F.~Beyer$^2$, P.~Indelicato$^4$, R.~M\"artin$^{2,5}$, R.~Reuschl$^{2,5}$ and Th.~St\"ohlker$^{2,6}$}

\address{$^1$Institut des Nanosciences de Paris, CNRS UMR7588 and Universit\'e Pierre et Marie Curie-Paris~6, Campus Boucicaut, 140 rue de Lourmel, F-75015 Paris, France}
\address{$^2$Gesellschaft f\"ur Schwerionenforschung, Planckstr. 1,  D-64291 Darmstadt, Germany}
\address{$^3$Nuclear Physics Division, Bhabha Atomic Research Centre, Mumbai - 400 085, India}
\address{$^4$Laboratoire Kastler Brossel, {\'E}cole Normale Sup{\'e}rieure; CNRS UMR8552; Universit{\'e} Pierre et Marie Curie-Paris6; Case 74, 4 Place Jussieu, F-75005  Paris, France}
\address{$^5$Institut f{\"u}r Kernphysik, Universit{\"a}t Frankfurt, D-60325 Frankfurt, Germany}
\address{$^6$Physikalisches Institut, Heidelberg, Philosophenweg 12, D-69120 Heidelberg, Germany}

\ead{martino.trassinelli@insp.jussieu.fr}

\begin{abstract}
We present the uncertainty discussion of a recent experiment performed at the GSI storage ring ESR for the accurate energy measurement of the  He-like uranium $1s2p\,^3P_2 \to 1s2s\, ^3S_1$ intra-shell transition. 
For this propose we used a Johann-type Bragg spectrometer that enables to obtain a relative energy measurement between the He-like uranium transition, about 4.51~keV, and a calibration x-ray source. 
As reference, we used the $K\alpha$ fluorescence lines of zinc and the Li-like uranium $1s^22p\, ^2\!P_{3/2} \to  1s^22s\, ^2\!S_{1/2}$ intra-shell transition from fast ions stored in the ESR. 
A comparison of the two different references, i.e., stationary and moving x-ray source, and a discussion of the experimental uncertainties is presented.  
\end{abstract}

\section{Introduction}
We present the uncertainty discussion of a recent experiment performed at the GSI (Darmstadt, Germany) for the accurate energy measurement of the  He-like uranium $1s2p\,^3P_2 \to 1s2s\, ^3S_1$ intra-shell transition. This measurement allows, for the first time, to test two-photons Quantum Electrodynamics in He-like heavy ions.
In this article we describe the techniques adopted in the measurement where x rays emitted from fast ions in a storage ring have been detected with a Bragg spectrometer. 
In particular, we study the contribution of the different uncertainties related to the relativistic velocity of the ions when a stationary or moving calibration x-ray source is considered. 
Additional details of such an experiment can be found in Ref.~\cite{Trassinelli2008}.

\section{Description of the set-up}
The experiment was performed at the GSI experimental storage ring ESR \cite{Franzke1987} in August 2007. Here, a H-like uranium beam
with up to $10^8$ ions was stored, cooled, and decelerated to an
energy of 43.57 MeV/u. Excited He-like ions were formed by electron
capture during the interaction of the ion beam with a supersonic
nitrogen gas-jet target. 
At the selected velocity,
electrons are primarily captured into shells with principal
quantum number of $n \leq 20$, which efficiently
populate the $n=2$ $^3\!P_2$ state via cascade feeding. 
This state decays to the $n=2$ $^3\!S_1$ state via an
electric dipole (E1) intra-shell transition (branching ration 30\%)
with the emission of photons of an energy close to 4.51~keV
detected by a Bragg spectrometer.

The crystal spectrometer \cite{Beyer1991} was mounted in the Johann geometry 
in a fixed angle configuration allowing for the detection of x rays with a Bragg angle $\Theta$ around $46.0^\circ$. 
The spectrometer was equipped with a Ge(220) crystal cylindrically bent, with a radius of curvature $R=800$~mm, and a newly fitted x-ray CCD camera (Andor DO420) as position sensitive detector.
The imaging properties of the
curved crystal were used to resolve spectral lines from fast x-ray
sources nearly as well as for stationary sources \cite{Beyer1988}.
For this purpose, it was necessary to place the Rowland-circle plane
of the spectrometer perpendicular to the ion--beam direction. 
For a minimizing the systematic effects due to the ion velocity and alignment uncertainties, the observation angle  $\theta = 90^\circ$ was chosen.
\begin{figure}[t!]
\includegraphics[trim= 5mm 6mm 0mm 8mm ,clip,width=\textwidth,height=0.25\textheight]{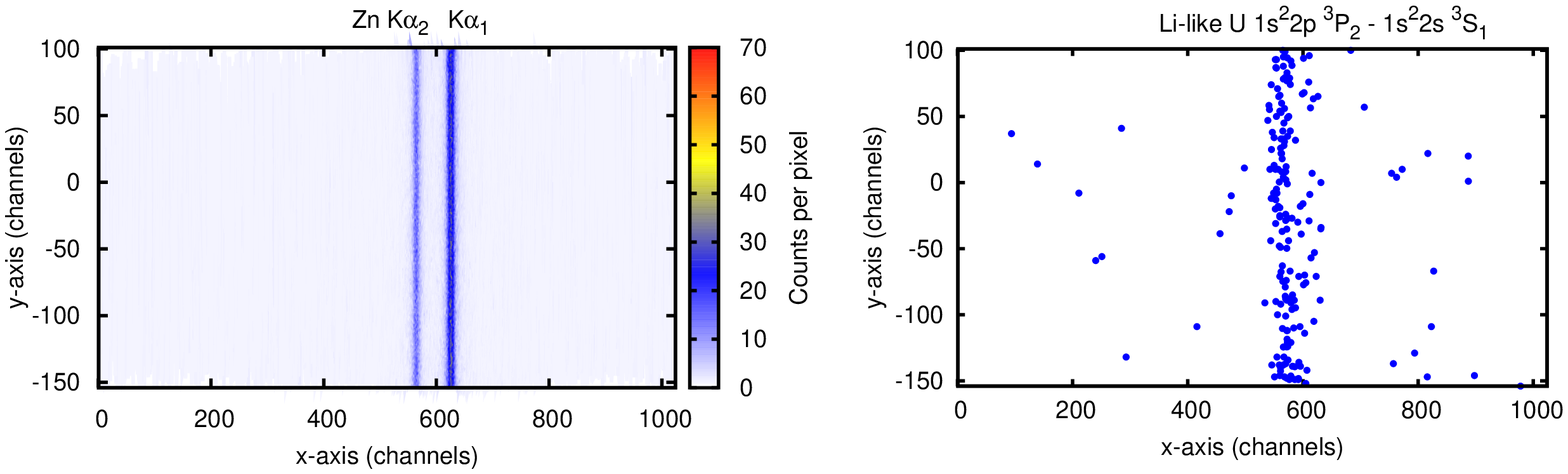}
\caption{\label{2d} Reflection of the zinc $K\alpha$ (left) and Li-like uranium intra-shell $1s^22p\, ^2\!P_{3/2} \to  1s^22s\, ^2\!S_{1/2}$  transition (right) on the Bragg spectrometer CCD. 
The transition energy increases with the increasing of x-position. 
The slightly negative slope of the line is due to the relativistic velocity of the Li-like ions.
}
\end{figure}

The value of the ion velocity $v$ was selected such that the photon energy, $E$ in the ion frame, was Doppler-shifted to the value $E_\text{lab} = 4.3$~keV in the laboratory frame, where $E_\text{lab} = E /[\gamma (1- \beta \cos \theta)]$, with $\beta = v / c$ ($c$ is speed of light) and  $\gamma = 1/\sqrt{1-\beta^2}$.
This value of $E_\text{lab}$ was chosen to have the He-like uranium spectral line position on the CCD close to the position of the 8.6 keV $K\alpha_{1,2}$ lines of zinc, which were observed in second order diffraction.
The zinc
lines were used for calibration and they were
produced by a commercial x-ray tube and a removable zinc plate
between the target chamber and the crystal. 
An image of the zinc $K\alpha$ lines from the Bragg spectrometer is presented in Fig.~\ref{2d} (left side). 

As an alternative method to the measurement of the He-like uranium intra-shell transition, we used
a calibration line
originating from fast ions, rather than one from the stationary
source. For this purpose the $1s^22p\, ^2\!P_{3/2} \to  1s^22s\, ^2\!S_{1/2}$
 transition in Li-like U at $4459.37 \pm 0.21$~eV \cite{Beiersdorfer1993,Beiersdorfer1995} was chosen. 
At the ESR, the
Li-like ions were obtained by electron capture into He-like
uranium ions. To match the energy of the He-like transition, an
energy of 32.63 MeV/u was used to Doppler-shift the Li-like
transition. 
An image of the Li-like uranium transition in the Bragg spectrometer is presented in Fig.~\ref{2d} (right side).

Starting from the Bragg's law in differential form, $\Delta E \approx
- E \, \Delta \Theta / \tan \Theta $, one
obtains an approximate dispersion formula that is valid for small
Bragg angle differences $\Delta \Theta$.
Taking into account the relativistic Doppler effect, the measured value of the He-like U transition is given by 
\begin{equation}
E = 
E_0 \frac n {n_0}
\frac {\left( 1 - \delta(E_0)/ \sin^2 \Theta_0 \right)}{\left( 1 - \delta(E)/ \sin^2 \Theta \right)} 
\cfrac {\gamma (1 - \beta \cos \theta )}{\gamma_0 (1 - \beta_0 \cos \theta )}
\left(
1 + \cfrac {\Delta x} {\tan \Theta_0 \ D}
\right),
\label{eq:bragg_rel}
\end{equation}
where $n_0$ and $n$ are the diffraction order of the He-like U and reference lines, respectively,  
$\Theta_0$ and $\Theta = \Theta_0 + \Delta \Theta$ the correspondent Bragg angle, $\Delta x$ is the position difference of the spectral lines on the CCD along the dispersion direction and $D$ is the crystal--CCD distance.
$\delta (E)$ is the deviation of the index of refraction $n_r(E) = 1 + \delta (E)$ of the crystal material from the unity, which depends on the energy $E$ of the reflected x ray ($\delta \sim  10^{-5}-10^{-6}$ typically).

In the case $n=n_0$, the corrections due to the refraction index and other energy dependent corrections for curved crystals \cite{Cembali1992,Chukhovskii1996} are negligible. 
In the case of a stationary calibration source, $\gamma_0=1$ and $\beta_0 = 0$.

The measured value of the He-like uranium transition energy  and additional information can be found in Ref.~\cite{Trassinelli2008}.
In the following section we present in detail the analysis of the systematic uncertainties.

\section{Evaluation of the experimental uncertainties}
One of the main sources of uncertainty in the present experiment is the low amount of collected data (see Fig.~\ref{2d}, right). This limits the accuracy of the He- and Li-like uranium line position, i.e., the accuracy of $\Delta x$, which is proportional to the energy uncertainty, $(\delta E)_\text{stat} \propto \delta (\Delta x)$ (see Eq.~\eqref{eq:bragg_rel}). In our specific experiment, characterized by the parameters listed in Table~\ref{tab:ex}, numerical values of $(\delta E)_\text{stat}$ are presented in Table~\ref{tab:err}.
Due to the high statistics in the stationary calibration source measurement, the Zn $K\alpha$ spectrum, $(\delta E)_\text{stat}$ is $\sim \sqrt{2}$ smaller than when the moving Li-like ion emission is used as a reference. 

In the case of systematic uncertainties, three major sources dominate: the accuracy of the reference energy $E_0$, of the ion velocity and of the observation angle $\theta$.
Similarly to the statistical uncertainty, the contribution of $\delta E_0$ is much smaller when Zn lines are used for calibration instead of the Li-like U transition (see Table~\ref{tab:err}). This is due to the high accuracy of the zinc $K\alpha$ transition energy, which in the case of $K\alpha_1$ is $8638.906 \pm 0.073$~eV \cite{Deslattes2003}, compared to the Li-like U transition accuracy of 0.21~eV \cite{Beiersdorfer1993,Beiersdorfer1995}.

If on one hand, Doppler tuning of the photon energy in the laboratory frame produce two important systematic uncertainty contributions, due to the ions velocity and the observation angle; on the other hand, it allows for detecting the different spectral lines in the same narrow spatial region of the CCD detector, i.e., $\Delta x / D \ll 1$. This results in a drastic reduction of other systematic effects  such as the influence of uncertainty of the crystal--CCD distance $D$, the accuracy of the CCD pixel size \cite{Indelicato2006}, the accuracy of the inter-plane distance of the crystal and effects from the optical aberrations in the Johann geometry set-up.
Systematic uncertainties related to the relativistic velocity of the ions are treated in detail in the following subsections.
\begin{table}[b!]
\begin{minipage}[b]{0.48\textwidth}\centering
\caption{\label{tab:ex}
Principal parameters of the ion beams for the He- and Li-like transition measurement (see text).}
\begin{tabular}{l | r r}
\br
 &He-like U&Li-like U\\
\mr
$\beta$ & 0.295578& 0.257944\\
$\gamma$ & 1.046771& 1.035026\\
$V$(Volt)& 23900& 17898 \\
$\theta$ & \multicolumn{2}{c}{$90.00^\circ  \pm 0.38^\circ$}\\
$a$(Volt)& \multicolumn{2}{c}{$0 \pm 10$}\\
$b$ & \multicolumn{2}{c}{$1. \pm 2 \times 10^{-4}$}\\
\br
\end{tabular}
\end{minipage}
\hspace{0.04\textwidth}
\begin{minipage}[b]{0.48\textwidth}\centering
\caption{\label{tab:err}
Different uncertainty contributions (in eV) when Li-like uranium and zinc transitions are used as the reference. }
\begin{tabular}{l | c c}
\br
 &Li-like U&Zn $K\alpha$\\
\mr
$(\delta E)_\text{stat}$ & 0.43& 0.30\\
$(\delta E)_{E_0}$& 0.21& 0.04\\
$(\delta E)_a$& $1 \times 10^{-3}$& 0.08\\
$(\delta E)_b$& 0.01& 0.04\\
$(\delta E)_\theta$& 0.11& 0.88\\
\mr
$(\delta E)_\text{TOT}$& 0.50& 0.93\\
\br
\end{tabular}
\end{minipage}
\end{table} 

\subsection{Ion velocity uncertainty}
The ion velocity in the storage ring is imposed by the velocity of the electrons in the electron cooler \cite{Franzke1987}. This is related to the cooler voltage $V$ by the simple relations
\begin{equation}
\gamma = 1 + \cfrac {e V} {m_e c^2}, \hspace{1cm}
\beta =  \cfrac {\sqrt{2 \frac {e V} {m_e c^2}+ \left( \frac {e V} {m_e c^2}\right)^2 }}{1 + \frac {e V} {m_e c^2}} \simeq \sqrt{2 \frac {e V} {m_e c^2}},
\end{equation}
where $m_e$ and $e$ are the mass and charge of the electron, respectively. The factor $e V / (m_e c^2)$ is in general very small, of the order of $4 \times 10^{-2}$ in our specific case.
The cooler voltage uncertainty $\delta V$ propagates to the energy uncertainty via the parameters $\gamma$ and $\beta$ in Eq.~\eqref{eq:bragg_rel}.
More precisely, $(\delta \gamma)_V  = e/(m_e c^2)\ \delta V$ and $(\delta \beta)_V \simeq [e /(2 m_e c^2 V)]^{1/2}\ \delta V $.
We note that $\delta \gamma / \delta \beta = \OO (\sqrt{e V / (m_e c^2)})$. For this reason an observation angle of $90^\circ$, where the effect of $\delta \beta$ is minimal, was chosen. In the following formulas we will consider only the case $\theta = 90^\circ$.


The uncertainty of the cooler voltage has two principal sources: the accuracy of the absolute value and its linearity.
The relation between the real voltage value $V_\text{real}$ and the set value $V$ can be written as $V= a + b\ V_\text{real}$, where the single uncertainties on the factors $a$ and $b$ has to be considered for our analysis.
With this notation, for an observation angle of  $\theta = 90^\circ$, the uncertainty propagation to the energy value is, for the case of a stationary calibration source,
\begin{equation}
\cfrac {(\delta E)_a} E  = \frac 1 \gamma  \left| \frac {\partial \gamma }{\partial a}\right| \ \delta a 
= \cfrac{e}{m_e c^2}\cfrac 1 \gamma   \ \delta a, \hspace{1cm}
\cfrac {(\delta E)_b} E  =  \frac 1 \gamma  \left| \frac {\partial \gamma }{\partial b}\right| \ \delta b  
= \cfrac{e V}{m_e c^2} \cfrac 1 \gamma \  \delta b, 
\end{equation}
and 
\begin{multline}
\cfrac {(\delta E)_a} E  = \frac {\gamma_0} \gamma \left| \frac {\partial}{\partial a} \left( \frac \gamma {\gamma_0} \right) \right| \ \delta a 
= \cfrac{e^2}{m_e^2 c^4}\cfrac {|V-V_0|} {\gamma\  \gamma_0} \ \delta a, \\
\cfrac {(\delta E)_b} E  = \frac {\gamma_0} \gamma \left| \frac {\partial}{\partial b} \left( \frac \gamma {\gamma_0} \right) \right| \ \delta b 
= \cfrac{e}{m_e c^2}  \cfrac {|V-V_0|} {\gamma\  \gamma_0} \  \delta b, 
\end{multline}
in the case of a moving calibration source, where $V_0$ and $\gamma_0$ are the corresponding parameters and where the approximation $V \approx V_\text{real}$ has been applied. 
A reduction of the uncertainties is obtained when the moving calibration source is used. 
The uncertainty $(\delta E)_a$ due to the offset error of $V$  is drastically decreased, by a factor $e|V-V_0|/(m_e c^2 \gamma_0) \ll 1$, here as the uncertainty due to the linearity is also reduced, but only by a factor $|V-V_0|/ (V \gamma_0)$. 
Numerical values for our experiments are given in Table~\ref{tab:ex} and \ref{tab:err}.

\subsection{Observation angle uncertainty}
If the effect of the uncertainty due to the ion velocity is minimal when $\theta = 90^\circ$, the effect of the uncertainty $\delta \theta$ of the observation angle itself is maximal.
In the case $\theta = 90^\circ$, starting from Eq.~\eqref{eq:bragg_rel}, for a stationary calibration source we have 
\begin{equation}
\cfrac {(\delta E)_\theta} E = \frac 1 {1- \beta \cos \theta} \left|  \frac d {d \theta} (1- \beta \cos \theta )\right|_{\theta = 90^\circ} \!\!\!  \delta \theta = \beta \delta \theta,
\end{equation} 
and
\begin{equation}
\cfrac {(\delta E)_\theta} E = \frac {1-\beta_0 \cos \theta} {1- \beta \cos \theta} \left|  \frac d {d \theta} \left(\frac {1- \beta \cos \theta }{1 - \beta_0 \cos \theta }\right) \right|_{\theta = 90^\circ} \!\!\!  \delta \theta 
= |\beta-\beta_0| \delta \theta.
\label{eq:oa}
\end{equation}
for a moving calibration source. 
Again, analogously to the calculation of the preceding subsection, a reduction of a factor $|\beta-\beta_0|/ \beta$ in the uncertainty is obtained when x rays from fast ions are used for the calibration.
In our experimental set-up, the value of $\delta \theta$ is principally due to the accuracy of the position of the gas-jet target with respect to the main axis of the spectrometer ($\pm 0.5$~mm).
Numerical values for our experiments are in Table~\ref{tab:ex} and \ref{tab:err}.
A direct evaluation of deviation $\Delta \theta$ from $90^\circ$ can be obtained via Eq.~\eqref{eq:oa} with the measurement of the Li-like uranium energy $4460.12 \pm 0.31$~eV (statistical uncertainty only) using the zinc $K\alpha$ lines as reference, and its comparison with the literature value $4459.37 \pm 0.21$~eV \cite{Beiersdorfer1993,Beiersdorfer1995}. We estimated $\Delta \theta = - 0.37^\circ \pm 0.18^\circ$ in agreement with the expected deviation (see Table~\ref{tab:ex}).

\section{Conclusions}
We present the uncertainty discussion of an accurate energy measurement on He-like uranium intra-shell transition obtained via Bragg spectroscopy of Doppler-tuned x rays emitted from fast ions. We have evaluated and compared the systematic uncertainties when a stationary or moving calibration source is used.  
In particular in our experiment, the use of the x-ray emission of fast Li-like uranium ions as reference enables to reduce systematics uncertainties by a factor of about 4.

\section*{References}

\end{document}